\documentclass[prd,aps,showpacs,tightenlines,nofootinbib,preprint,preprintnumbers]{revtex4}
\usepackage{graphics}
\usepackage{amsfonts}
\usepackage{amssymb}
\usepackage{latexsym}

\newcommand{\bear}{\begin{array}}  \newcommand{\eear}{\end{array}}
\newcommand{\bea}{\begin{eqnarray}}  \newcommand{\eea}{\end{eqnarray}}
\newcommand{\beq}{\begin{equation}}  \newcommand{\eeq}{\end{equation}}
\newcommand{\bef}{\begin{figure}}  \newcommand{\eef}{\end{figure}}
\newcommand{\bec}{\begin{center}}  \newcommand{\eec}{\end{center}}
  
\newcommand{\lmk}{\left(}  \newcommand{\rmk}{\right)}
\newcommand{\lkk}{\left[}  \newcommand{\rkk}{\right]}

\newcommand{\bib}{\bibitem}

\def\IBB#1#2#3{{\bf #1}, #2 (20#3)}
\def\IBID#1#2#3{{\it ibid}. {\bf #1}, #2 (19#3)}
\def\IBIDD#1#2#3{{\it ibid}. {\bf #1}, #2 (20#3)}

\def\CQGG#1#2#3{Class. Quantum Grav. {\bf #1}, #2 (20#3)}

\def\JHEPP#1#2#3{J. High Energy Phys. {\bf #1}, #2 (20#3)}

\def\JP#1#2#3{J. Phys. A {\bf #1}, #2 (19#3)}

\def\NPB#1#2#3{Nucl. Phys. {\bf B#1}, #2 (19#3)}

\def\PLB#1#2#3{Phys. Lett. B {\bf #1}, #2 (19#3)}
\def\PLBB#1#2#3{Phys. Lett. B {\bf #1}, #2 (20#3)}

\def\PRD#1#2#3{Phys. Rev. D {\bf #1}, #2 (19#3)}
\def\PRDD#1#2#3{Phys. Rev. D {\bf #1}, #2 (20#3)}

\def\PRL#1#2#3{Phys. Rev. Lett. {\bf#1}, #2 (19#3)}
\def\PRLL#1#2#3{Phys. Rev. Lett. {\bf#1}, #2 (20#3)}

\def\PTP#1#2#3{Prog. Theor. Phys. {\bf #1}, #2 (19#3)}

\begin{document}

\thispagestyle{empty}
{\baselineskip0pt
\leftline{\large\baselineskip16pt\sl\vbox to0pt{\hbox{\it Department of
Mathematics and Physics}
               \hbox{\it Osaka City  University}\vss}}
\rightline{\large\baselineskip16pt\rm\vbox to20pt{\hbox{OCU-PHYS-229}
            \hbox{AP-GR-24}
\vss}}%
}
\vskip3cm

\title{Black Strings in Our World}


\author{Hiroyuki Nakano${}^{1}$, Ken-ichi Nakao${}^{1}$, and Masahide
Yamaguchi${}^{2}$}
\affiliation{${}^{1}$Department of Mathematics and Physics, 
Graduate School of Science,
Osaka City University, Osaka 558-8585, Japan \\
${}^{2}$
Department of Physics and Mathematics, Aoyama Gakuin
University, Sagamihara 229-8558, Japan}


\date{\today}

\begin{abstract}
The brane world scenario is a new approach to resolve the problem on how
to compactify the higher dimensional spacetime to our 4-dimensional
world. One of the remarkable features of this scenario is the higher
dimensional effects in classical gravitational interactions at short
distances.  Due to this feature, there are black string solutions in our
4-dimensional world.  In this paper, assuming the simplest model of
complex minimally coupled scalar field with the local $U(1)$ symmetry,
we show a possibility of black-string formation by merging processes of
type I long cosmic strings in our 4-dimensional world. No fine tuning for the
parameters in the model might be necessary.  

\end{abstract}


\maketitle

\section{Introduction}

The brane-world scenario is a model inspired by superstring theory which
is one of the most promising theories for the unification of all
elementary interactions including gravity.  The basic idea of this
scenario has been given by Arkani-Hamed, Dimopoulos, and Dvali; the
matter and gauge fields except for gravitational field are confined within the
brane which is a 4-dimensional timelike submanifold in the higher
dimensional spacetime\cite{ref:ADD-braneI,ref:ADD-braneII}.  Then
putting their basic idea to account, Randall and Sundrum gave very
original two models for compactification of higher dimensional spacetime
to our 4-dimensional world\cite{ref:RS-braneI,ref:RS-braneII}.  
In this paper, we assume RS model II in which there is a 
4-dimensional positive tension
brane embedded in 5-dimensional spacetime called
bulk\cite{ref:RS-braneII}.  In order to describe the bulk dimension, we
adopt Gaussian normal coordinate $w$ which is set so that the brane is
located at $w=0$.  Then in the framework of this model, the spacetime
geometry is determined by the following 5-dimensional Einstein
equations,
\begin{equation}
G_{ab}={8\pi \over M_{\rm pl}{}^{3}}\left(T_{ab}-\sigma h_{ab}\right)
\delta(w)-\Lambda_5 g_{ab},
\end{equation} 
where $g_{ab}$, $G_{ab}$, $M_{\rm pl}$, $\Lambda_5$ are the 5-dimensional
metric tensor, Einstein tensor, Planck mass and cosmological constant,
respectively, $\sigma$ is the tension of the brane, 
$T_{ab}$ is the stress-energy tensor of the matter or
gauge fields confined within the brane, $h_{ab}$ is the intrinsic
metric of the brane, $\delta(w)$ is the Dirac's delta function, 
and we have adopted the natural unit $c=\hbar=1$. In
accordance with the Randall and Sundrum, we assume the model of $8\pi
\sigma/M_{\rm pl}{}^3=\sqrt{-6\Lambda_5}$, where $\Lambda_5$ is assumed
to be negative.  This model recovers the 4-dimensional Einstein gravity
for long range gravitational interaction within the brane.

The 5-dimensional Planck mass $M_{\rm pl}$ is related to the
4-dimensional one $m_{\rm pl}~(=1.2\times10^{19}{\rm GeV})$ by $M_{\rm
pl}{}^3=m_{\rm pl}{}^2/l$, where $l=\sqrt{-6/\Lambda_5}$ is the $AdS$
length scale.  The experimental constraint on this $AdS$ length is
$l<0.1$mm\cite{ref:Newton-Correction}.  
Therefore 5-dimensional Planck mass $M_{\rm pl}$ is then
written as
\begin{equation} 
M_{\rm pl}= 6.7\times 10^8~l_{0.1}^{~-\frac{1}{3}}~{\rm GeV},
\label{eq:M5-value}
\end{equation}
where we introduce a normalized $AdS$ length defined by 
\begin{equation}
l_{0.1}=\frac{l}{0.1{\rm mm}}.
\end{equation}
The brane tension $\sigma$ is written as
\begin{equation}
\sigma={3M_{\rm pl}{}^3\over 4\pi l}
\simeq
         \lmk 3.4 \times 10^3~
          l_{0.1}^{~-\frac{1}{2}} ~ {\rm GeV} 
         \rmk^4.
\label{eq:sigma-def}
\end{equation}
It should be noted that the brane tension $\sigma$ is much smaller than
5-dimensional Planck scale $M_{\rm pl}{}^4$ if $l$ is of order
$0.1{\rm mm}$.

In the brane world scenario, quantum effects in gravity will appear at
the energy scale over 5-dimensional Planck mass $M_{\rm pl}$ which may
be much smaller than 4-dimensional Planck mass $m_{\rm pl}$. On the
other hand, effects of the higher dimension appear in the gravity at the
distance scale shorter than the $AdS$ length scale $l$.
Since $l$ is much larger than $M_{\rm pl}{}^{-1}$ in the case of 
$l\simeq0.1$mm, the higher dimensional effects can be
important even in the classical gravitational interactions.  One of
remarkable features of the higher dimensional Einstein gravity is the
existence of black-string solutions which do not exist in 
4-dimensional Einstein theory of gravity with the dominant energy
conditions on matter fields\cite{ref:Hayward}.  Therefore if the distribution of
mass-energy is infinitely long but sufficiently thin, a black string
forms also in the RS model II\cite{ref:NNM}.  This is consistent with
the extended version of hoop conjecture\cite{ref:IN}.

One of the present authors, Nakamura and Mishima showed that a black
string forms in the RS model II 
if the circumferential radius, i.e., ``thickness'' of the
cylindrically distributed mass within the brane is smaller than its
gravitational thickness $r_{\rm g}$ defined by
\begin{equation}
r_{\rm g}:={\mu \over M_{\rm pl}{}^3}, 
\label{eq:rg-definition}
\end{equation}
as long as $\mu/m_{\rm pl}{}^2 \ll 1$, where $\mu$ is the line energy
density, i.e., the energy per unit length scale; the inequality
$\mu/m_{\rm pl}{}^2\ll 1$ is equivalent to the condition that the
gravity produced by the brane tension $\sigma$ is much smaller than that
of the black string at the horizon.  However, at first glance, it is not
likely that such configurations are realized in our universe, even if
the brane world scenario describes our universe.  An exception might be
a cosmic string and thus in this paper, we focus on it and study the
possibility of its gravitational collapse to form a black string. 

This paper is organized as follows. In Sec.II, we briefly review the 
local $U(1)$ gauged cosmic string and give a crude criterion for 
the formation of a black string by the gravitational collapse of 
a cosmic string. In Sec.III, we show the scenario 
of the black-string formation caused by mergers of type I cosmic 
strings in the expanding universe.  
Finally, Sec.IV is devoted to summary and discussion about several issues 
of black strings in the expanding universe. 

\section{Gravitational Collapse of Type I Cosmic String}


We assume the $U(1)$ gauged cosmic string \cite{VS} whose motion is
confined within the brane. The Lagrangian density of the complex scalar
field $\phi$ and $U(1)$ gauge field $A^{\mu}$ is
\begin{equation}
{\cal L}=-h^{\mu\nu}
\left(\partial_\mu+ieA_\mu\right){\bar\phi}
\left(\partial_\nu-ieA_\nu\right)\phi+{1\over4}F^{\mu\nu}F_{\mu\nu}
+V\left(|\phi|\right), 
\end{equation}
where $e$ is a positive real constant, 
$\partial_{\mu}$ is the partial derivative operator within the
brane, $F_{\mu\nu}=\partial_\mu A_\nu-\partial_\nu A_\mu$ is 
the field strength tensor, and the potential $V(|\phi|)$ is assumed to be
\begin{equation} 
V\left(|\phi|\right)={1\over4}\lambda\left(|\phi|^2-\eta^2\right)^2,
\end{equation} 
where $\lambda$ and $\eta$ are positive real parameters.

In this model, there are two length scales; one is Compton length of the
scalar field $r_{\rm s}$ and the other is that of the gauge field
$r_{\rm v}$.  These are, respectively, given as
\begin{equation}
r_{\rm s}= {1\over \sqrt{\lambda}\eta}~~~~{\rm and}~~~~
r_{\rm v}={1\over \sqrt{2}e\eta}.
\label{eq:scales}
\end{equation}
Here we introduce a parameter $q$ defined by
\begin{equation}
q:={2e^2\over \lambda}.
\end{equation}
The model of $q>1$ ($r_{\rm s}>r_{\rm v}$) is called type I, that of
$q=1$ ($r_{\rm s}=r_{\rm v}$) is called Bogomol'nyi limit and the case
of $0<q<1$ ($r_{\rm s}<r_{\rm v}$) is called type II. The line energy
density $\mu_{\rm sl}$ of a single winding cosmic string has been
numerically computed and its fitting formula is given
as\cite{ref:Hog-Prim}
\begin{equation}
\mu_{\rm sl}={1.04\pi\over(2q^2)^{0.195}}~\eta^2
=:\alpha(q)\eta^2. 
\label{eq:sl-line-density}
\end{equation} 
The above formula is valid to about 5$\%$ accuracy in the range
$0.01<2q^2<100$. Although this formula is derived by assuming no
self-gravity, this might be sufficient for our purpose to get a crude
criterion for black-string formation.

First we consider black-string formation by a cosmic string of single
winding number.  The thickness of a type I cosmic string is given by
$r_{\rm s}$, while that of Bogomol'nyi limit or of type II cosmic string
is equal to $r_{\rm v}$ Hence the type I cosmic string forms into a black
string if $r_{\rm s}\lesssim r_{\rm g}$, while in the case of
Bogomol'nyi limit and type II cosmic string, the condition on
black-string formation is given by $r_{\rm v}\lesssim r_{\rm g}$.  By
Eqs.(\ref{eq:rg-definition}), (\ref{eq:scales}) and
(\ref{eq:sl-line-density}), these conditions lead to
\begin{equation}
\lmk \frac{\eta}{M_{\rm pl}} \rmk \gtrsim 
{1 \over \omega^{1/6}\alpha^{1/3}}
\label{eq:eta-criterion}
\end{equation}
with
\begin{equation}
\omega=
\left\{\begin{array}{lll}
       \lambda &~~~~ \mbox{for type I}, \\
        2e^2 &~~~~ \mbox{for Bogomol'nyi limit or type II}.
       \end{array}\right. \label{eq:p-def}
\end{equation}
Therefore in order that a single winding cosmic string collapses to form into 
a black string, the phase transition should occur at the energy scale
almost equal to or larger than 5-dimensional Planck scale $M_{\rm pl}$
if $\lambda$ and $2e^2$ are of order unity.  The phase transition over
Planck scale seems to be unnatural.  The parameters $\lambda$ for type I
and $2e^2$ for Bogomol'nyi limit or type II should be larger than $10^6$
so that the breaking scale $\eta$ is sufficiently lower than
5-dimensional Planck scale $M_{\rm pl}$.  However, on the physical
ground, it is preferable that dimensionless parameters in the physical
model are of order unity. Therefore in the following discussion, we
assume moderate values for $\lambda$ and $2e^2$ and do not consider
black-string formation by a single winding cosmic string.

Large line energy density and small thickness of a cosmic string is
necessary for black-string formation by its gravitational
collapse. Hence multiple winding cosmic strings will be important. In
fact, the line energy density of a multiple winding cosmic string is
larger than that of a single winding one while its thickness is 
almost the same as that of a single winding one\cite{VS,ref:dLTV}.  
The multiple winding
cosmic strings in type I model are stable while those of Bogomol'nyi
limit or type II models are unstable and will decay into single winding
ones. Therefore in the following discussion, we focus on type I cosmic
strings.  The line energy density $\mu_{\rm ml}(n)$ of a cosmic string
with the winding number $n$ might be written in the form,
\begin{equation}
\mu_{\rm ml}(n)=\beta(n)n \mu_{\rm sl}=\alpha(q)\beta(n) n\eta^2,
\label{eq:ml-line-density-1}
\end{equation} 
where $\beta(n)$ represents the effect of the interaction energy and
should be less than unity due to the stable nature of the type I
multiple winding cosmic string. Then, in the case of a multiple winding cosmic
string, the condition for black-string formation is given by
\beq
  n > n_{\rm c} \equiv \frac{1}{\alpha \beta \sqrt{\lambda}} 
           \lmk \frac{M_{\rm pl}}{\eta} \rmk^3.
\label{eq:ml-condition}
\eeq
If the winding number $n$ of a cosmic string is larger than the critical value 
$n_{\rm c}$ at its formation, it would collapse and a black string would 
be formed. However, the production probability of multiple winding
cosmic strings by Kibble mechanism is very small\cite{ON}. Hence we
investigate the possibility of mergers of type I long cosmic strings in the
expanding universe as a formation process of a highly multiple winding
cosmic string.

\section{Formation of Black Strings by Merger of Cosmic Strings}

Bettencourt and Kibble discussed the merger of type I long strings and
analytically estimated escape velocities for various intersection angles
and the model parameter $q$\cite{ref:Betten-Kibble}.  The numerical
simulations were performed by Bettencourt, Laguna and
Matzner\cite{ref:Betteb-Lag-Matz} and their results are basically
consistent to the previous analytic estimation. For example, in the case
of $q>2$, the escape velocity between the cosmic strings of intersection
angle $30^{\circ}-60^\circ$ is about 0.1. Since each collision will be
highly inelastic, the merger can occur even if the colliding velocity is
larger than 0.1. According to the numerical simulations of cosmological
evolution of type II cosmic strings, typical velocity of cosmic string
is $0.63-0.67$ during radiation dominated era and $0.57-0.65$ during
matter dominated era \cite{AT,BB,MSM}. However, in these simulations,
peculiar nature of type I cosmic strings is not considered
adequately. Thus, the above results are not necessarily applicable to
type I cosmic strings. Furthermore, it is pointed out that the estimated
velocity includes not only coherent velocity but also that due to small
scale motion so that coherent velocity is much smaller and could be 0.15
\cite{VV}. This means that the merging probability is not so
small. Therefore, the winding number might grow as a result of merger,
though the string network basically grows according to the scaling
solution because loops are also produced through intercommutation. Then,
we will investigate the development of a typical winding number in the
context of scaling evolution of the cosmic string network, in which the
typical network of cosmic strings grows with the cosmic
time.\footnote{Such scaling property is confirmed not only for local
strings \cite{Kibble,AT,BB,AS} but also for global strings \cite{gs}
and global monopoles \cite{gm}.}

The homogeneous and isotropic universe model in the framework of the
RS-model II is considered here. Assuming the bulk is the exact
5-dimensional $AdS$ spacetime, the equation for the scale factor $a$ of
the Friedmann brane is given by
\begin{eqnarray}
\left({{\dot a}\over a}\right)^2
=\left({4\pi\over3M_{\rm pl}{}^3}\right)^2\rho(2\sigma+\rho)
-{k\over a^2}+{\Lambda_4\over3},
\end{eqnarray}
where the dot means the time derivative, $\rho$ is the energy density of
the matter and radiation, $k$ is the sign of the spatial curvature and
$\Lambda_4$ is the 4-dimensional cosmological constant
\cite{ref:Ida}. The $\rho$-square term represents the higher dimensional
effect and thus if $\rho$ is larger than the brane tension $\sigma$ but
smaller than the 5-dimensional Planck scale $M_{\rm pl}{}^4$, the higher
dimensional classical gravity is realized. In the present setting,
$\sigma$ is given by Eq.(\ref{eq:sigma-def}) and thus, 
when the temperature of the universe $T$ is much higher than
$\sigma^{1/4}$, the $\rho$-square term becomes dominant. On the other
hand, it is negligible when $T \ll \sigma^{1/4}$.

Before the cosmic string network obeys the scaling solution, the
friction coming from interactions with surrounding plasma
dominates its dynamics. The transition temperature $T_{\rm tr}$ from the
friction dominated era to the scaling regime is determined by the
balance between the Hubble parameter and the friction coefficient
$T_{\rm tr}^3 /\nu \mu$, where $\nu$ is a numerical factor related 
to the coupling between the cosmic string and surrounding plasma and 
takes a value of $1-10^3$. Assuming that the $\rho$-square term is negligible
and single winding strings are dominant, the transition temperature is
given and constrained from above, as follows. 
\beq
  T_{\rm tr} \simeq \frac{\nu\mu_{\rm sl}}{m_{\rm pl}} \ll 
             \frac{\nu M_{\rm pl}^2}{m_{\rm pl}} 
                   \simeq 3.6 \times 10^{-2} ~
             \frac{\nu}{l_{0.1}^{~\frac{2}{3}}} ~{\rm GeV}.
\eeq 
Thus, unless $l$ is too small, $T_{\rm tr}$ is much lower than
$\sigma^{1/4}$. Then, we assume that the $\rho$-square term is
subdominant at the transition time.

In the scaling regime, the energy density of the long cosmic strings is given
by
\beq
  \rho_{\infty} = \frac{\mu_{\rm ml}(n)}{L^2},
\eeq 
where $L$ is a scale characterizing the cosmic string network and is
proportional to the cosmic time in the manner $L = \gamma t$ with 
a positive constant $\gamma$. In this regime, mean coherent velocity $v$ of 
long cosmic strings is constant. 
Then denoting the merging probability per collision by $p$, 
the evolution equation for the number $N$ of mergers per the characteristic 
length scale $L$ is given by
\beq
  \frac{dN}{dt} = \frac{p v}{L}(N+1)+\frac{Z}{L}\frac{dL}{dt}N, 
\label{eq:evolution}
\eeq
where $Z$ is non-negative parameter smaller than or equal to 
unity. The first term of R.H.S. in the above equation comes from 
the increase of merger number due to collisions, while the second 
term describes the evolution of configurations of cosmic strings 
already interconnected to each other. 
A merger of two cosmic strings at $t=t_{\rm m}$ might produce 
a segment interconnected 
to each other over the characteristic length scale $\gamma t_{\rm m}$ 
at the moment of this merger. 
There is large uncertainty on the evolution of the 
interconnected segment. If the length of interconnection does not 
change from $\gamma t_{\rm m}$ even for $t>t_{\rm m}$, 
the parameter $Z$ should vanish. 
In this case, the length scale over which $N$ cosmic strings are 
interconnected to each other is the characteristic length 
$\gamma t_{1}$, where $t=t_{1}$ is the moment of the first merger. 
By contrast, if the interconnected segment evolves 
in the manner of a zipper\cite{ref:Betten-Kibble,ref:Betteb-Lag-Matz}, 
all of the cosmic strings attached to the interval of the characteristic 
length $\gamma t$ of some cosmic string might completely merge 
into a cosmic string over the length scale $\gamma t$. 
In this case, the parameter $Z$ should be taken unity. 

We assume that the probability $p$ is constant. Then the solution of  
Eq.(\ref{eq:evolution}) is given by 
\begin{equation}
N=\frac{P}{P+Z}\left[\left(1+\frac{P+Z}{P}N_{\rm tr}\right)
\left(\frac{t}{t_{\rm tr}}\right)^{P+Z}-1\right],
\label{eq:number}
\end{equation}
where $P:=pv/\gamma$ and $N_{\rm tr}$ is an integration constant 
which corresponds to the number of mergers at the transition time 
$t=t_{\rm tr}$. 
$N_{\rm tr}$ may take a large value because a typical
coherent velocity is smaller in the friction dominated epoch so that the
merger may happen more often. However, the evolution of the cosmic
string network in the friction dominated era strongly depends on its
distribution at its formation. So, we do not mention this any more. Instead,
we simply assume that $N_{\rm tr}$ vanishes to give a conservative
estimate because the main purpose of the present paper is to show that
black strings can be actually produced in our universe.

Since the winding number $n$ randomly increases by each merger because a
string encounters a string or an anti-string with equal probability, it
is estimated as $n\simeq N^{1\over2}$.  When a winding number exceeds
the critical value $n_c$, black strings are formed. Assuming the
formation time $t_{\rm f}$ is much larger than the transition time
$t_{\rm tr}$, $t_{\rm f}$ is estimated from Eqs. (\ref{eq:ml-condition})
and (\ref{eq:number}) as
\beq
  t_{\rm f} = t_{\rm tr} \lkk \frac{P+Z}{P\alpha^2 \beta^2 \lambda}
                      \lmk \frac{M_{\rm pl}}{\eta} \rmk^6 
                     \rkk^{\frac{1}{P+Z}}.
\eeq
At the formation, the ratio of the energy density of black strings
$\rho_{\rm BS}$ to the total energy density of the universe $\rho_{\rm
tot}$ is estimated as
\bea
  \frac{\rho_{\rm BS}}{\rho_{\rm tot}}(t_{\rm f})
     &=& \lmk \frac{n_{c} \alpha \beta \eta^2}{\gamma^2 t_{\rm f}^{~2}} \rmk
       \lmk \frac{3m_{\rm pl}^2\epsilon^2}{8\pi t_{\rm f}^{~2}} \rmk^{-1}
=2.5\times 10^{-20}\frac{M_{\rm pl}}
{\epsilon^{2}\gamma^{2}\lambda^{\frac{1}{2}}
l_{0.1}^{~\frac{2}{3}}\eta}
\eea
with $\epsilon = 1/2$ for the case of the formation in 
the radiation dominant era and $\epsilon
= 2/3$ for the formation in the matter dominant era. 
In the following estimations, we assume the cosmic concordance model 
of density parameter of non-relativistic matter 
$\Omega_{\rm M}\simeq0.3$ and that of the 
cosmological constant $\Omega_{\Lambda}\simeq0.7$ 
since recent observational results are consistent with this 
model. Therefore the formation of black strings 
in the cosmological constant dominant era is possible 
but we do not consider it here. 

In the present scenario, the winding number  
is not constant along a cosmic string. 
In the case of non-zipper evolution, the 
winding number varies on the length scale $\gamma t_{\rm 1}$, 
while in the case of complete zipper, its variation scale evolves 
as $\gamma t$. Then segments with sufficiently large winding number 
will be enclosed by horizons which are compact and disconnected 
to each other. Therefore, exactly speaking, the formed black objects 
are not strings but rather thin worms of finite lengths. However, 
once a black worm  is formed on a segment of large winding number, it 
will grow to indefinitely long configuration by swallowing up the cosmic 
strings interconnected to this black worm and hence it might 
as well be called a black string. 

Black strings are formed in the radiation dominated era when 
$\eta\geq \eta_{\rm RD}$, where
\bea
\frac{\eta_{\rm RD}}{M_{\rm pl}} \equiv 
\left(8.8\times10^{-16}\right)^{\frac{P+Z}{4(P+Z)+6}}
{\cal F}
\label{eq:eta-RD}
\eea
with 
\bea
{\cal F} \equiv 
\left[
\left(
\frac{l_{0.1}^{~~\frac{4}{3}}}{{\cal N}_{\rm tr}^{~\frac{1}{2}}\nu^{2}}
\right)^{P+Z}
\frac{P+Z}{\alpha^{2(P+Z+1)}
\beta^{2}\lambda P}
\right]^{\frac{1}{4(P+Z)+6}}.
\eea
Here, ${\cal N}_{\rm tr}$ is a factor coming from the effective number of
distinct helicity states at the transition time.  The above condition
has been obtained from the condition that $t_{\rm f}$ is smaller than or
equal to the matter-radiation equality time $t_{\rm
eq}\simeq(16\pi^3T_{\rm eq}^{~4}/(45m_{\rm pl}^{~2}))^{-1/2}/2$ and we
have used $T_{\rm eq} \simeq 0.76$~eV.  On the other hand, the condition
that black strings are formed by now is obtained from the condition
$t_{\rm f}\leq t_0$ which leads to $\eta\geq \eta_{\rm MD}$, where
\bea
\frac{\eta_{\rm MD}}{M_{\rm pl}}\equiv
\left(
4.2\times 10^{-21}
\times\frac{14{\rm Gyr}}{t_0}
\right)^{\frac{P+Z}{4(P+Z)+6}}
{\cal F}
\label{eq:eta-MD}
\eea
and $t_0$ is the present cosmic time. Note 
that $\eta_{\rm RD}$ and $\eta_{\rm MD}$ are smaller as $l$ is smaller.
Here we have used the relation 
$t_{\rm tr}
=(4\pi^3 {\cal N}_{\rm tr}T_{\rm tr}^{~4}/(45m_{\rm pl}^{~2}))^{-1/2}/2$.

\subsection{Non-Zipper Evolution}

First we consider the case of non-zipper evolution $Z=0$.
Though there are various unknown physics, 
we assume the merger probability $p$, typical coherent velocity $v$ and 
the parameter $\gamma$ determining characteristic length $L$ so that
\begin{equation}
P=0.4 \lmk \frac{p}{0.8} \rmk
      \lmk \frac{v}{0.15} \rmk
      \lmk \frac{0.3}{\gamma} \rmk.
\end{equation}
Substituting this value into Eqs.(\ref{eq:eta-RD}) and (\ref{eq:eta-MD}), 
we obtain 
\begin{eqnarray}
\frac{\eta_{\rm RD}}{M_{\rm pl}}&=&0.16 {\cal F}_{\rm nz}, \\
\frac{\eta_{\rm MD}}{M_{\rm pl}}&=&8.5\times10^{-2} 
\left(\frac{14{\rm Gyr}}{t_0}\right)^{\frac{1}{19}}{\cal F}_{\rm nz}, 
\end{eqnarray}
where
\begin{equation}
{\cal F}_{\rm nz}\equiv l_{0.1}^{~\frac{4}{57}}
{\cal N}_{\rm tr}^{~-\frac{1}{38}}
\nu^{-\frac{2}{19}}
\alpha^{-\frac{7}{19}}\beta^{-\frac{5}{19}}\lambda^{-\frac{5}{38}}.
\end{equation}
Note that although $\eta_{\rm RD}$ and $\eta_{\rm MD}$ are smaller as $l$ 
is smaller, their dependence on $l$ is very weak. It is easily seen 
from the above equation that 
the dependence of $\eta_{\rm RD}$ and $\eta_{\rm MD}$ on $\alpha$ 
is the strongest while their dependence on the other 
unknown parameters ${\cal N}_{\rm tr}$, $\nu$, $\beta$ and $\lambda$ 
is very weak. Therefore if $\alpha$ is order of unity, 
we may say that if the merger probability 
$p$ is sufficiently large, the formation of black strings is possible even though 
the breaking scale $\eta$ is much smaller than the 5-dimensional Planck scale. 

\subsection{Zipper Evolution}

In the case of the zipper evolution, we assume that $P$ is much smaller 
than unity for conservative estimation. Then we find 
\begin{eqnarray}
\frac{\eta_{\rm RD}}{M_{\rm pl}}&=&3.1\times10^{-2} 
{\cal F}_{\rm zp}, \\
\frac{\eta_{\rm MD}}{M_{\rm pl}}&=&9.2\times10^{-3} 
\left(\frac{14{\rm Gyr}}{t_0}\right)^{\frac{1}{10}}{\cal F}_{\rm zp}, 
\end{eqnarray}
where
\begin{equation}
{\cal F}_{\rm zp}\equiv l_{0.1}^{~\frac{2}{15}}
{\cal N}_{\rm tr}^{~-\frac{1}{20}}
\nu^{-\frac{1}{5}}
\alpha^{-\frac{2}{5}}\beta^{-\frac{1}{5}}\lambda^{-\frac{1}{10}}P^{-\frac{1}{10}}.
\end{equation}
Also in this case, the dependence of $\eta_{\rm RD}$ and $\eta_{\rm MD}$ on 
$\alpha$ is the strongest. Here it should be noted that the dependence of 
$\eta_{\rm RD}$ and $\eta_{\rm MD}$ on $P$ is very weak. This means that even though  
the probability $p$ of a merger is very small, large winding number of 
a cosmic string might be realized and as a result, the black string 
might form even for rather small breaking scale $\eta$ if the  
interconnected segments sufficiently rapidly evolve in the manner of zipper. 

\subsection{Present Energy Density of Black Strings}

Once black strings are formed, reconnection and formation of loops
cease. Therefore, the energy density of black strings is not suppressed
in the same manner as cosmic strings, instead its energy density might
decrease in proportional to the inverse squared of the scale factor.
Then, the present energy density of black
strings is estimated as 
\bea
  \frac{\rho_{\rm BS}}{\rho_{\rm tot}}(t_0)
     = \frac{\rho_{\rm BS}}{\rho_{\rm tot}}(t_f) 
     \left(\frac{a(t_{\rm eq})}{a(t_f)}\right)^{\frac{2-2\epsilon}{\epsilon}}
     (1+z_{\rm eq})(1+z_\Lambda)^{-3},
\eea
where $z_{\rm eq}$ is the redshift of the matter-radiation equality time, 
which is estimated to be $3233$ \cite{WMAP}, while $z_\Lambda$ is the 
redshift of matter-$\Lambda$ equality time estimated to be $0.33$. 
Therefore in the case of the black-string formation  
in the radiation dominant era, the 
present energy density of the black string is given by
\begin{eqnarray}
  \frac{\rho_{\rm BS}}{\rho_{\rm tot}}(t_0)
&=&0.16~{\cal N}_{\rm tr}^{~\frac{1}{2}}
l_{0.1}^{~-2}
\nu^2 \gamma^{-2} \nonumber \\
&\times&\left[
\alpha^{2(P+Z+1)}\beta^2 \lambda^{\frac{2-(P+Z)}{2}}
\left(\frac{\eta}{M_{\rm pl}}\right)^{3(P+Z+2)}
\left(\frac{P}{P+Z}\right)
\right]^{\frac{1}{(P+Z)}},
\end{eqnarray}
while in the case of the black-string formation in the matter dominant era, it is 
given by
\begin{eqnarray}
  \frac{\rho_{\rm BS}}{\rho_{\rm tot}}(t_0)
&=&8.4\times 10^{-7}~{\cal N}_{\rm tr}^{~\frac{1}{3}}
l_{0.1}^{~-\frac{14}{9}}
\nu^{\frac{4}{3}} \gamma^{-2} \nonumber \\
&\times&\left[
\alpha^{4(P+Z+1)}\beta^4 \lambda^{\frac{4-3(P+Z)}{2}}
\left(\frac{\eta}{M_{\rm pl}}\right)^{5(P+Z)+12}
\left(\frac{P}{P+Z}\right)^2
\right]^{\frac{1}{3(P+Z)}}. 
\end{eqnarray}
Thus, since $\eta < M_{\rm pl}$, and since $P+Z$ is at most 
order of unity, the present energy density of black
strings is subdominant as long as the parameters $\alpha$, $\beta$, and
$\lambda$ are also order of unity and $l_{0.1}$ is not too small. 
However we should note that in contrast to $\eta_{\rm RD}$ and $\eta_{\rm MD}$, 
the energy density $\rho_{\rm BS}$ strongly depends on these 
unknown parameters. Thus it is possible that the black strings are 
a dominant component in the present universe. 

\section{Summary and Discussion}

In this paper, we discuss the black-string formation by considering
multiple winding strings. Though the formation probability of such
strings is small at the formation of strings, they can be formed through
merging processes of type I cosmic strings. Based on several numerical
simulations of type I cosmic strings, we discuss the formation of
multiple winding cosmic strings and subsequent collapse into black
strings. We show that black strings can be formed in our 4-dimensional
world for reasonable sets of parameters.  However, in the numerical
simulations of type I cosmic strings, the setting of collisions of two
strings is given by hand and far from realistic. Collisions of type I
cosmic strings will be investigated for the realistic evolution in the
expanding universe. 

A black string is a source of gravitational radiation.  The
shortest wavelength of the gravitational waves might be determined by
Gregory-Laflamme instability\cite{ref:GL-instability}.  The
Schwarzschild black string solution is unstable against linear
perturbations of the wavelength longer than several times of 
$r_{\rm g}$. In the present scenario, the gravitational thickness $r_{\rm g}$ 
cannot exceed the $AdS$ length $l$. If $r_{\rm g}>l$, the gravity 
near the horizon of the black string is 
that for the 4-dimensional spacetime. This fact leads to an inconsistency 
since there is no black string solution in 4-dimensional gravity. 
Thus black strings might emit gravitational waves of various wavelength larger 
than some very short lower cutoff which must be less than $l$. 

Like as a cosmic string,  self-intersections of a black string will occur. 
In this case, black loops are produced but those might be kept attaching to 
their parent black string and then the loop structures may disappear by
``shrinking''. If the black loops leave their parent black string, 
naked singularities appear. At present, it is not clear whether isolated 
loops are released from their parent black string. 
However, we may say that such a process will produce much gravitational waves. 
The intersections between individual black strings may also produce 
very complicated network and much gravitational radiation. 
The spectrum and amplitude of such a gravitational radiation 
should be studied in detail and this is a future work. 

It is worthwhile to note that the condition of black-string formation is 
identical to that for the topological inflation\cite{ref:T-inflation}. 
However, in the present case, the deflation rather than the inflation 
may be realized. Farhi and Guth 
presented a proof for the impossibility of inflation in the 
laboratory\cite{ref:FG}. 
Although the present situation is not identical to the assumption  
in their proof, the same result might be obtained. Therefore, inside the 
black-string horizon, the multiple winding cosmic string might form spacetime 
singularities rather than the child universes\cite{ref:KSato}.   

The evaporation of the black string due to Hawking effect is also 
an important problem. Here we should note that the black string produced 
by the multiple winding cosmic string also has a winding number which is 
observable for outside observers and is a topological invariant. 
Thus in order that the black string with the non-vanishing 
winding number evaporates completely, the winding number should be 
released. This means that cosmic strings with some winding numbers 
should be created outside the horizon  
through the quantum processes in order for the 
complete evaporation. However the creation 
probability of infinite cosmic strings seems to be very small and therefore 
the black string considered in this paper might not evaporate completely, 
although detailed study is necessary. 

\section*{Acknowledgements}

We are grateful to H.~Ishihara and colleagues in the astrophysics and 
gravity group of Osaka City University for their useful and helpful 
discussion and criticism. KN would like to thank M. Nagasawa for giving 
very useful information about the physical process of cosmic strings 
and also thank J. Soda for his useful suggestion on the instability of 
black strings. This work is supported by the Grant-in-Aid for Scientific 
Research (No.16540264) and that for Young Scientists (No.5919), from JSPS.


\begin{thebibliography}{99}

\bibitem{ref:ADD-braneI}
N.~Arkani-Hamed, S.~Dimopoulos, and G.~Dvali, 
\PLB{429}{263}{98}.
\bibitem{ref:ADD-braneII}
I.~Antoniadis, N.~Arkani-Hamed, S.~Dimopoulos, and G.~Dvali,
\PLB{436}{257}{98}.
\bibitem{ref:RS-braneI}
L.~Randall and R.~Sundrum, 
\PRL{83}{3370}{99}.
\bibitem{ref:RS-braneII}
L.~Randall and R.~Sundrum, 
\PRL{83}{4690}{99}.
\bibitem{ref:Newton-Correction}
G. C.~Long, H.~Chang, and J.~Price, 
\NPB{529}{23}{99};
C. D. Hoyle {\it et al.},
\PRLL{86}{1418}{01}.
\bibitem{ref:Hayward}
S. A. Hayward, 
\CQGG{17}{1749}{00}.
\bibitem{ref:NNM}
K.~Nakao, K.~Nakamura, and T.~Mishima, 
\PLBB{564}{143}{03}.
\bibitem{ref:IN}
D.~Ida and K.~Nakao, 
\PRDD{66}{064026}{02}.
\bib{VS}
For review, see A. Vilenkin and E. P. S. Shellard, 
{\it Cosmic String and Other Topological Defects}
(Cambridge University Press, Cambridge, England, 1994).
\bibitem{ref:Hog-Prim}
H. M.~Hodge and J. R.~Primack, 
\PRD{43}{3155}{91}.
\bibitem{ref:dLTV}
A. A.~de Laix, M.~Trodden and T.~Vachaspati, 
\PRD{57}{7186}{98}.
\bib{ON}
T. Okabe and M. Nagasawa, 
\PLB{416}{49}{99}.
\bibitem{ref:Betten-Kibble}
L. M. A. Bettencourt and T. W. B. Kibble, 
\PLB{332}{297}{94}.
\bibitem{ref:Betteb-Lag-Matz}
L. M. A. Bettencourt, P.~Laguna, and R. A.~Matzner, 
\PRL{78}{2066}{97}.
\bib{AT}
A. Albrecht and N. Turok,
\PRD{40}{973}{89}
\bib{BB}
D. P. Bennett and F. R. Bouchet,
\PRD{41}{2408}{90}.
\bib{MSM}
J. N. Moore, E. P. S. Shellard, and C. J. A. P. Martins,
\PRDD{65}{023503}{01}.
\bib{VV}
T. Vachaspati and A. Vilenkin,
\PRL{67}{1057}{91}.
\bib{Kibble}
T. W. B. Kibble, 
\JP{9}{1387}{76}.
\bib{AS}
B. Allen and E. P. S. Shellard,
\PRL{64}{119}{90}.
\bib{gs}
M. Yamaguchi, J. Yokoyama, M. Kawasaki,
\PTP{100}{535}{98};
M. Yamaguchi, M. Kawasaki, and J. Yokoyama,
\PRL{82}{4578}{99};
M. Yamaguchi, 
\PRD{60}{103511}{99};
M. Yamaguchi, J. Yokoyama, and M. Kawasaki,
\IBIDD{61}{061301(R)}{00};
M. Yamaguchi and J. Yokoyama,
\IBIDD{66}{121303(R)}{02};\IBB{67}{103514}{03}.
\bib{gm}
M. Barriola and A. Vilenkin,
\PRL{63}{341}{89};
D. P. Bennett and S. H. Rhie,
\IBID{65}{1709}{90};
U. Pen, D. N. Spergel, and N. Turok,
\PRD{49}{692}{94};
M. Yamaguchi,
\PRDD{64}{081301(R)}{01}; \IBB{65}{063518}{02}.
\bibitem{ref:Ida}
D.~Ida, 
\JHEPP{09}{014}{00}.
\bib{WMAP}
C.~L.~Bennett {\it et al.},
Astrophys.\ J.\ Suppl. {\bf 148}, 1 (2003).
\bibitem{ref:GL-instability}
R. Gregory and R. Laflamme, 
\PRL{70}{2837}{93}.
\bibitem{ref:T-inflation}
A. Linde, \PLB{327}{208}{94};
A. Vilenkin, \PRL{72}{3137}{94}.
\bibitem{ref:FG}
E. Farhi and A. H. Guth, 
\PLB{183}{149}{87}
\bibitem{ref:KSato}
K. Sato, M. Sasaki, H. Kodama and K. Maeda,
\PTP{65}{1443}{81}.

\end{thebibliography}
\end{document}